\documentclass[journal]{IEEEtran}

\usepackage{amsmath}
\usepackage{graphicx}
\usepackage{tikz}
\usepackage{color}
\usepackage{amsmath}
\usepackage{subfig}
\usepackage{amssymb}
\usepackage{balance}
\usepackage{math}
\graphicspath{{./Fig/}}

\usepackage[final]{review}

\begin{document}


\title{ Audio-noise Power Spectral Density  Estimation Using Long Short-term Memory }
%

\author{Xiaofei Li, Simon Leglaive, Laurent Girin and Radu Horaud

 \thanks{Xiaofei Li, Simon Leglaive and Radu Horaud are with Inria and with Univ. Grenoble Alpes,, France. 
 E-mail: \texttt{first.last@inria.fr} 
 }
 
 \thanks{Laurent Girin is with Inria and with Univ. Grenoble Alpes, Grenoble-INP, GIPSA-lab, France. 
 E-mail: \texttt{laurent.girin@grenoble-inp.fr}
 }
 
 }

%
%
%

%
\maketitle
\begin{abstract}

We propose a method using a long short-term memory (LSTM) network to estimate the noise power spectral density (PSD) of single-channel audio signals represented in the short time Fourier transform (STFT) domain. An LSTM network common to all frequency bands is trained, which processes each frequency band individually by mapping the noisy STFT magnitude sequence  to its corresponding noise PSD sequence. Unlike deep-learning-based speech enhancement methods that learn the full-band spectral structure of speech segments, the proposed method exploits the sub-band STFT magnitude evolution of noise with a long time dependency, in the spirit of the unsupervised noise estimators described in the literature. Speaker- and speech-independent experiments with different types of noise show that the proposed method outperforms the unsupervised estimators, and generalizes well to noise types that are not present in the training set.

\end{abstract}

\section{Introduction}
\addnote[motivation]{1}{Noise power spectral density (PSD) estimation is a prerequisite for many audio applications, such as 
speech enhancement \cite{ephraim1984,cohen2001,erkelens2007}, voice activity detection \cite{sohn1999,li2016iwaenc}, acoustic environment identification \cite{malik2013} and noise-aware training of speech enhancement network \cite{xu2014,fu2016}, to cite a few}.  \addnote[latency1]{1}{Noise PSD estimation is generally performed  in the time-frequency (TF) domain, and  in an online manner}.
The local minimum of the smoothed noisy signal periodogram, searched in a sliding window, is widely employed for noise PSD estimation \cite{martin2001}. Due to the spectral sparsity of speech, the local minimum point is assumed to locate in a speech absence segment, and thus it corresponds to a noise-only segment. The local minimum is multiplied with a compensation factor leading to noise PSD estimate in the minimum statistics algorithm \cite{martin2001}. Based on the local minimum, the improved minima controlled recursive averaging algorithm (MCRA) \cite{cohen2003} first estimates the speech presence probability (SPP) for each frame, and then averages the noisy signal periodogram weighted by SPP. In \cite{rangachari2006}, a non-linear averaging of the past spectral values is proposed to track the local minimum, which circumvents the possible tracking latency when applying the minimum-search window. 
Instead of using the local minimum, other regional statistics such as normalized variance and median crossing rate are used to estimate the SPP in \cite{li2016icassp}, which also circumvents the aforementionned possible tracking latency. The minimum mean-squared error (MMSE) based methods \cite{hendriks2010,gerkmann2012} estimate the noise PSD by recursively averaging the posterior mean of the noise periodogram given the noisy speech periodogram, which can be interpreted as a voice activity detector. In the MMSE-based methods, the required parameters of the probablistic model, i.e. noise and speech PSDs, are approximated by their estimates at the previous frame.  
The above mentioned methods, i.e. \cite{martin2001,cohen2003,rangachari2006,li2016icassp,hendriks2010,gerkmann2012}, are all unsupervised and applied separately for each frequency bin. They explicitly or implicitly detect the noise-only segments, and estimate the noise PSD during these segments. To do that, they exploit the difference in noise and speech characteristics, i.e. noise is assumed to be more stationary than speech, and the speech TF representation is assumed to be more sparse than the noise one. 
Therefore, these methods are suitable for reasonably non-stationary background noise, but not for highly non-stationary (transient) noise, i.e. noise with a PSD that can vary suddenly. 

Recently, supervised deep-learning-based speech enhancement has been largely investigated, see \cite{wang2018} for an overview. These methods use a neural network to map noisy speech features to clean speech features. The input features, e.g. cepstral coefficient and linear prediction based features, generally encode the full-band structure of  noisy speech spectra. The output target vector generally consists of either the clean speech STFT magnitude vector or an ideal binary (or ratio) mask vector to be applied on the corresponding noisy speech STFT frame.  Widely-used speech enhancement neural networks include feed-forward neural network (FNN) and recurrent neural network (RNN). The temporal dynamics of speech can be modeled  by stacking context frames in the FNN input, while it is automatically modeled by RNN. The ideal binary (ratio) mask can be considered as an SPP estimate. Therefore, it can be further used for noise variance (or covariance for the multichannel case) estimation \cite{heymann2016,higuchi2016,xiao2017,zhang2017,boeddeker2018}. In \cite{weninger2014}, an LSTM RNN is employed to estimate log-Mel spectrograms not only for clean speech, but also for noise. In \cite{papadopoulos2017}, instead of estimating the noise PSD for each TF bin, the global signal-to-noise-ratio (SNR) of a long-term noisy speech signal is estimated using an FNN. 

In this work, we propose an online method for estimating the noise PSD individually at each frequency band, thus following the same principle as the unsupervised methods \cite{martin2001,cohen2003,rangachari2006,li2016icassp,hendriks2010,gerkmann2012}, but leveraging an LSTM RNN. Such a network is able to efficiently model the temporal dynamics of audio signal \cite{weninger2014,chen2017}. In the STFT domain, a sequence of noisy speech STFT magnitudes within a small subband (3 consecutive frequency bins) is input to the LSTM network, which outputs the corresponding sequence of noise (log-scale) PSD estimate at the corresponding central frequency bin.
This process is applied for all frequencies with the same unique LSTM network. The network is expected not only to learn a regression function from the input sequence to the output sequence, but also to learn to extract low-level information such as the local minimum \cite{martin2001,cohen2003,rangachari2006}, regional statistics \cite{li2016icassp} and signal correlation between neighboring frequency bins, and to automatically implement mid-level information processing that are useful for noise PSD estimation, such as the SPP calculation \cite{cohen2003} and the recursive update process \cite{hendriks2010,gerkmann2012}.  Compared with deep-learning-based speech enhancement methods that learn the full-band spectral structure, the proposed method is expected to have better generalization capabilities. 
Indeed, the proposed LSTM network does not rely on the full-band spectral structure, and thus has to model much smaller variability with respect to speakers, speech content (including different languages) and noise types. In addition, due to the small feature dimension and variability, the proposed method requires a smaller network, and thus less training data and a lower computation cost at both training and prediction time. However, the proposed method mainly relies on the speech/noise discriminative information, typical of unsupervised methods.
It is thus poorly suitable for transient noises with abrupt variations.


\section{Noise PSD Estimation with LSTM Network}
\label{sec:method}


We consider a single-channel signal in the STFT domain:
\begin{equation}
x(k,l)=s(k,l)+u(k,l),
\end{equation}
where $x(k,l)$, $s(k,l)$ and $u(k,l)$ are the (complex-valued) STFT coefficients of the microphone, speech and noise signals, respectively, $k=0,\cdots,K-1$ and $l$ are the frequency and frame indices, respectively. The speech signal $s(k,l)$ and noise signal $u(k,l)$ are assumed to be independent random variables. The noise PSD is defined as $\lambda_u(k,l)=\mathbb{E}[|u(k,l)|^2]$, where $\mathbb{E}[\cdot]$ and $|\cdot|$ denote expectation and modulus, respectively.
In this work, we consider a ``reasonably'' non-stationary background noise with slowly-varying PSD.
Therefore, the noise PSD at a given frame can be approximately calculated by averaging the noise periodogram over a small number of adjacent past frames. For online calculation, recursive averaging is used: $\lambda_u(k,l)=\alpha{\lambda}_u(k,l-1)+(1-\alpha)|u(k,l)|^2$,
where $\alpha$ is the smoothing factor. However, the true noise signal is usually unoberseved and the goal of noise PSD estimation is to compute $\lambda_u(k,l)$ from the observed noisy speech signal $x(k,l)$. In this work, we employ LSTM for this aim. 

\subsection{Input Feature}
Unsupervised methods \cite{martin2001,cohen2003,rangachari2006,li2016icassp,hendriks2010,gerkmann2012} only rely on local information provided by the sequence of noisy speech STFT magnitude coefficients, considering each frequency bin independently. The phase information is  ignored since it does not carry information about the noise PSD. In this work, we also would like to exploit the signal correlation between  neighboring frequency bins. Thence, for frequency $k$, the STFT magnitude vector 
\begin{equation}
\mathbf{x}(k,l) = [|x(k-1,l)|,|x(k,l)|,|x(k+1,l)|]^{\top}
\end{equation} 
is used as the input feature to the LSTM network, where $^\top$ denotes vector transpose. Note that, for $k=0$ or $K-1$, the non-existing neighbour data is replaced with data at frequency $k$, which is thus duplicated.
 For frame $l$, to perform the online estimation, we take the current and previous frames
\begin{equation}\label{eq:xseq}
\mathcal{X}(k,l)=  \big(\mathbf{x}(k,l-T+1),\dots,\mathbf{x}(k,l) \big),
\end{equation} 
as the input sequence, where $T$ is the sequence length. To facilitate the network training, the input sequence has to be normalized to equalize the input level. Based on some pilot experiments, the mean of the sequence at frequency bin $k$, i.e. $\mu(k,l) = \frac{1}{T}\sum_{l'=l-T+1}^l |x(k,l')|$, is used for normalization. The input sequence it thus finally given by:  
\begin{equation}\label{eq:xseqnor}
\tilde{\mathcal{X}}(k,l) = \mathcal{X}(k,l)/\mu(k,l).
\end{equation}



\subsection{Output Target} 
For frequency $k$ at frame $l$, the ground truth noise PSD sequence
\begin{equation}
\Lambda_u(k,l)=\big(\lambda_u(k,l-T+1),\dots,\lambda_u(k,l)\big), 
\end{equation} 
is taken as the target. According to the input sequence normalization, we use $\mu^2(k,l)$ to normalize the noise PSD sequence. Finally, the logarithm of the normalized sequence, i.e.
\begin{equation}\label{eq:yseqnor}
\tilde{{\Lambda}}_u(k,l) = \text{log}(\Lambda_u(k,l)/\mu^2(k,l))
\end{equation}
is taken as the output target sequence. 
During test, the predicted output $\hat{\tilde{\Lambda}}_u(k,l)$ is transformed back to the original domain as $\hat{{\Lambda}}_u(k,l)) = e^{\hat{\tilde{{\Lambda}}}_u(k,l)}\mu^2(k,l)$, which is the noise PSD estimation for TF bin $(k,l)$. 

\subsection{Noise PSD Estimation Network}

RNN transmits the hidden units along time step. To avoid the problem of exponential weight decay (or explosion) along time steps, LSTM introduces an extra memory cell, which conveys the information along time step respectively to the hidden units. The memory cell allows to learn long-term dependencies. For the detailed structure of LSTM, see the seminal paper \cite{hochreiter1997}.

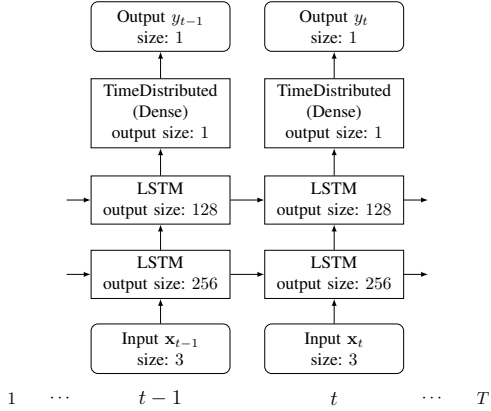
\begin{figure}[t!]
\centering

\pgfdeclarelayer{background}

\pgfsetlayers{background}

\usetikzlibrary{shapes.geometric,backgrounds,arrows,calc,fit}
\tikzstyle{io} = [rectangle, rounded corners, minimum width=2.8cm, minimum height=1.0cm, text centered, draw=black,align=center]
\tikzstyle{lstm} = [rectangle,  minimum width=2.8cm, minimum height=0.5cm, text centered, draw=black,align=center]
\tikzstyle{data1} = [rectangle, rounded corners, minimum width=2.4cm, minimum height=0.8cm, text centered, draw=black,align=center]
\tikzstyle{nobox} = [rectangle, rounded corners, draw=none,fill=white,minimum width=1cm, minimum height=0cm, text centered, align=center]
 \tikzstyle{constant} = [rectangle, minimum width=3.5cm, minimum height=1.1cm,text centered, draw=black,align=center] 
 \tikzstyle{arrow} = [->,>=latex]

\resizebox{0.75\columnwidth}{!}{%
\begin{tikzpicture}


\node (xt) [io] {{Input} $\mathbf{x}_t$ \\ size: $3$};

\node (t) [below of=xt,node distance=1.0cm] {\large $t$};
\node (d1) [right of=t,node distance=2cm] {$\cdots$};
\node (T) [right of=d1,node distance=1.0cm] {$T$};

\node (l1t) [lstm,above of =xt,node distance=1.5cm] {LSTM \\ output size: $256$};

\node (l2t) [lstm,above of =l1t,node distance=1.5cm] {LSTM \\ output size: $128$};

\node (denset) [lstm,above of =l2t,node distance=1.75cm] {TimeDistributed \\ (Dense) \\ output size: $1$};

\node (yt) [io,above of =denset,node distance=1.75cm] {{Output} $y_t$ \\ size: $1$};

\node (xtm1) [io,left of =xt,node distance=3.5cm] {{Input} $\mathbf{x}_{t-1}$ \\ size: $3$};

\node (tm1) [below of=xtm1,node distance=1.0cm] {\large $t-1$};
\node (d2) [left of=tm1,node distance=2cm] {$\cdots$};
\node (t1) [left of=d2,node distance=1.0cm] {$1$};

\node (l1tm1) [lstm,above of =xtm1,node distance=1.5cm] {LSTM \\ output size: $256$};

\node (l2tm1) [lstm,above of =l1tm1,node distance=1.5cm] {LSTM \\ output size: $128$};

\node (densetm1) [lstm,above of =l2tm1,node distance=1.75cm] {TimeDistributed \\ (Dense) \\ output size: $1$};

\node (ytm1) [io,above of =densetm1,node distance=1.75cm] {{Output} $y_{t-1}$ \\ size: $1$};

\draw [arrow] (xt) --  (l1t);
\draw [arrow] (l1t) --  (l2t);
\draw [arrow] (l2t) --  (denset);
\draw [arrow] (denset) --  (yt);

\draw [arrow] (xtm1) --  (l1tm1);
\draw [arrow] (l1tm1) --  (l2tm1);
\draw [arrow] (l2tm1) --  (densetm1);
\draw [arrow] (densetm1) --  (ytm1);

\draw [arrow] (l1tm1) --  (l1t);
\draw [arrow] (l2tm1) --  (l2t);

\node (l2tm2) [nobox, left of =l2tm1,node distance=2.4cm] {};
\draw [arrow]  (l2tm2) --  (l2tm1);
\node (l1tm2) [nobox, left of =l1tm1,node distance=2.4cm] {};
\draw [arrow]  (l1tm2) --  (l1tm1);
\node (l2tp1) [nobox, right of =l2t,node distance=2.4cm] {};
\draw [arrow]  (l2t) --  (l2tp1);
\node (l1tp1) [nobox, right of =l1t,node distance=2.4cm] {};
\draw [arrow]  (l1t) --  (l1tp1);

\end{tikzpicture}
}
\vspace{-0.0cm}
\caption{Diagram of the proposed network.} 
\label{fig:lstm}
\vspace{-0.0cm}
\end{figure}

Fig. \ref{fig:lstm} shows the diagram of the network used in this work, where two LSTM layers are stacked. 
The output vector of the second LSTM layer is transformed to the output target, i.e.  the noise PSD estimate, through a time-distributed dense layer. The time-distributed dense layer shares the parameters for all the time steps. 
The whole system has about 0.46 M learnable parameters.
Note that the input sequence $\mathbf{x}_t, \ t=1,\dots,T$ and output sequence  $y_t, \ t=1,\dots,T$ represent one sequence defined by (\ref{eq:xseqnor}) and (\ref{eq:yseqnor}), respectively, with any frequency index $k$ and frame index $l$.

\section{Experiments}
\label{sec:experiment}

\subsection{Dataset and data pre-processing}

Twelve types of noise from the NOISEX92 database \cite{varga1993} were used: 
white, babble, pink, buccaneer1, buccaneer2, f16, hfchannel,  factory1, factory2, destroyerengine, destroyerops, m109.  We used clean speech signals from the TIMIT database \cite{garofolo1988}. Each noise signal was split into three sections used for training (70\%), validation (10\%) and test (20\%), respectively, which means different noise instances are used for training and test. Speech signals from the TIMIT training set were used for training, and \addnote[split]{1}{speech signals from the TIMIT \emph{Diverse} test set were equally split without speaker overlap, and were used for validation and test, respectively}.  This means that the experiments are both speaker-independent and speech-content-independent. All signals are resampled to $16$~kHz. 

To generate noisy signals, speech and noise signals were randomly selected from their corresponding train/validation/test set, and mixed with a given SNR. The noisy signal and pure noise signal were transformed to the STFT domain using a 512-sample (32~ms) Hamming window with a frame step of 256 samples. After pilot experiments, the training sequence length was set to $T=128$ frames (about 2~s). 
Four SNRs were used to create the training data, namely $\{-3, 3, 9, 15\}$~dB. For each type of noise and each SNR, $500$ seconds of noisy data were generated. For training, we picked one pair of input/output sequences (\ref{eq:xseqnor}) and (\ref{eq:yseqnor}) every $64$ frames from the training data, which makes two consecutive sequences being not highly similar and guarantees high variability of training sequences. 
Four SNRs were used to create the validation and test data, i.e. $\{0, 5, 10, 15\}$~dB. For each type of noise and each SNR, $45$ and $90$ seconds of noisy data were generated for validation and test, respectively.

\subsection{Network Training}

Remind that in principle one single LSTM network is designed to process all frequency bins and all types of noise. Therefore, all training sequences (with different frequency bin $k$ from $0$ to $K-1$, different $l$ index, and different speech content, noise types and SNRs) were presented to the same network. \addnote[noisetype]{1}{However, in practice, two networks were trained: The first one, referred to as LSTM-12, uses all twelve noise types. The second one, referred to as LSTM-9, excludes three of them, namely pink, buccaneer2 and factory2. 
By comparing the performance measures of these two networks on pink, buccaneer2 and factory2, we can evaluate the generalization ability of the proposed network in terms of noise type}.  For these two networks, a total of $500\times 4\times 12/3600\approx 6.7$ hours and $500\times 4\times 9/3600= 5$ hours of signal were used for training, respectively, from which a total of about $5.8\times 10^6$ and $4.3\times 10^6$ training sequences were generated, respectively. These sequences were shuffled during training. 

\addnote[training]{1}{The mean squared error (MSE) was used as the training cost. We used the Keras framework \cite{chollet2015keras}  to implement the proposed method. The Adam optimizer \cite{kingma2014adam} was used with a learning rate of 0.001. The batch size was 512. The training process was early-stopped with a patience of two epochs}.

\subsection{Noise PSD Prediction Setting}
\addnote[bptt]{1}{The networks were trained with a sequence length set to $T=128$, in other words, the back propagation through time \cite{werbos1990} goes through $128$ time steps}.  At test time, even though the length of test sequence is not theoretically constrained to be $128$, this choice leads to the best performances in pilot experiments. 
To process a long test signal, a sliding window is applied to form the successive test sequences with length of 128 frames. 
For one test sequence, the prediction error decreases with the increasing of time step, since more past information is used by the late time steps. 
Therefore, to achieve the smallest prediction error, only the prediction of the last time step should be output as the estimated noise PSD for the corresponding frame. \addnote[latency2]{1}{For this case, the moving step of the sliding window is set to one, and the noise PSD of one frame is estimated using this frame and its previous 127 frames, in other words, there is no estimation latency}.  However, the computation cost of this scheme is very high, since one sequence is processed to obtain the noise PSD estimation for a single frame. \addnote[latency3]{1}{In our experiments, to reduce the computation cost, we output the prediction of the last $32$ time steps as the estimated noise PSD for the corresponding frames, and thus the moving step for the sliding window is set to 32 frames. Note that this leads to an estimation latency of $32$ frames}.

\begin{figure}[t]
\centering
\includegraphics[width=.99\columnwidth]{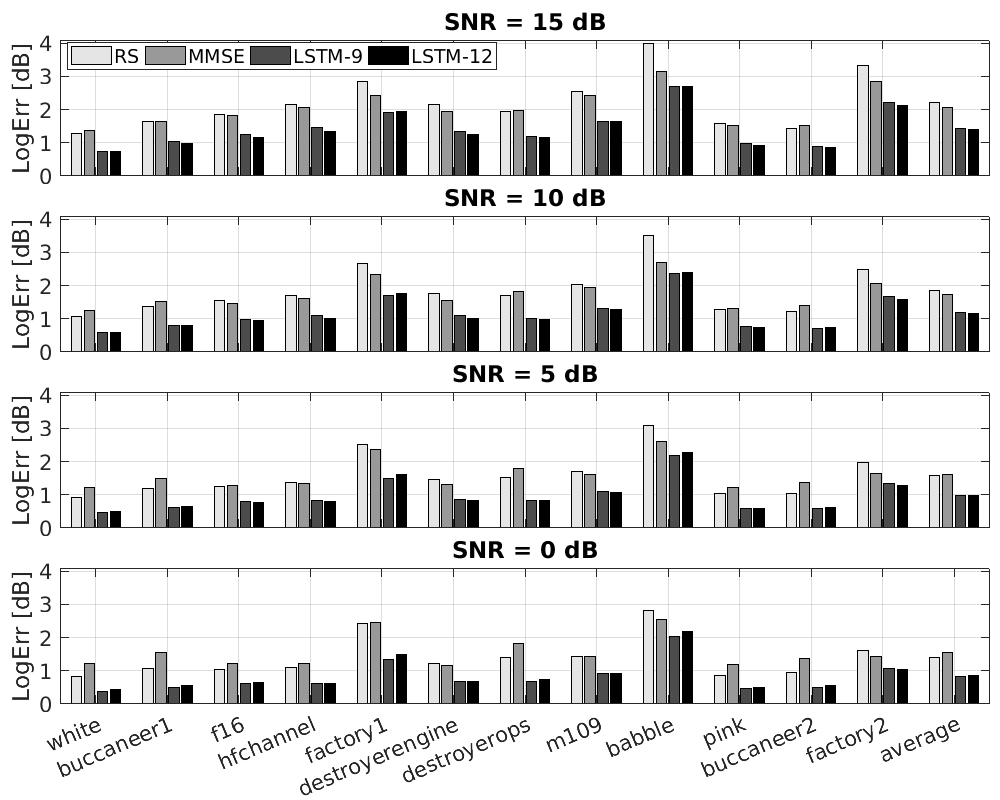}
\vspace{-4mm}
\caption{\small{Logarithmic error of noise PSD estimation. }}
\label{fig:logerr}
\end{figure}

\begin{figure}[t]
\centering
\vspace{-2mm}
\includegraphics[width=.9\columnwidth]{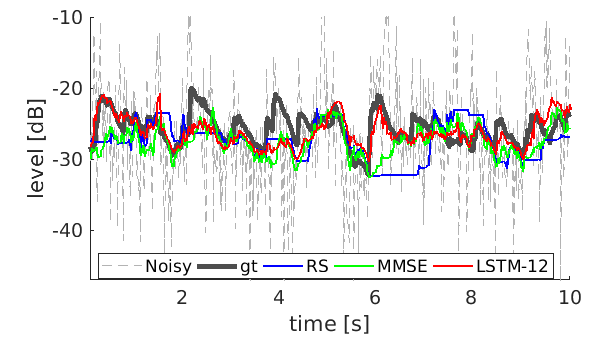}
\vspace{-2mm}
\caption{\small{An example of audio-noise PSD estimation at 3 kHz. Noise signal is factory1, and SNR is 10 dB. `gt' means ground truth.}}
\label{fig:example}
\vspace{-2mm}
\end{figure}

\subsection{Experimental Results}

Two unsupervised methods are used as baselines: the regional statistics (RS) method of \cite{li2016icassp} and the MMSE-based method of \cite{gerkmann2012}.
The symmetric segmental logarithmic error (LogErr, in dB) \cite{hendriks2008} is taken as the criterion for evaluating the noise PSD estimation performance. The smaller LogErr is, the better the estimation.  
The estimated noise PSD is used to derive the optimally-modified log-spectral amplitude estimator \cite{cohen2001} for speech denoising. The perceptual evaluation of speech quality (PESQ) \cite{rix2001} and segmental SNR (SNR$_\text{seg}$, in dB) \cite{gerkmann2012} are applied on the resulting denoised signal to evaluate the denoising performance.  Note that SNR$_\text{seg}$ is different from the SNR mentioned above. The latter is computed using the power of the entire signal, while the former is computed by averaging the SNR over the signal segments. The signal segments are set to have a length of $10$ ms and zero overlap, and noise-only frames are excluded for the calculating of SNR$_\text{seg}$. For both PESQ and SNR$_\text{seg}$, the higher the better.

\begin{figure}[t]
\centering
\includegraphics[width=.75\columnwidth]{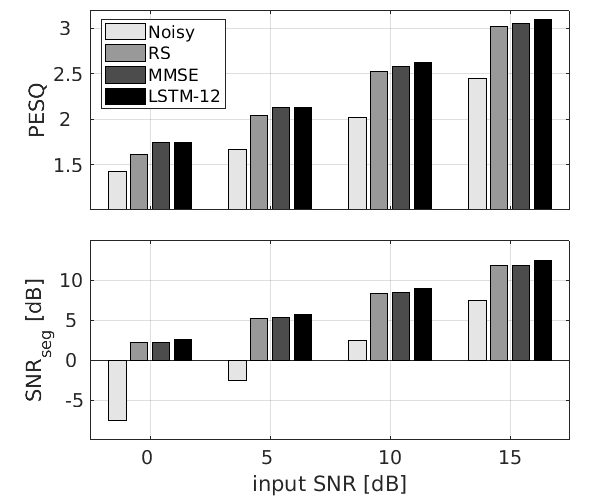}
\vspace{-0mm}
\caption{\small{PESQ and SNR$_\text{seg}$ scores averaged over all noise types.}}
\label{fig:se}
\vspace{-2mm}
\end{figure}

\subsubsection{Noise PSD Estimation Results}
Fig. \ref{fig:logerr} shows the LogErr values obtained for the different types of noise. 
It can be seen that the proposed LSTM-based method significantly outperforms the two unsupervised baseline methods for all SNRs and all noise types. This shows the superiority of the data-driven supervised method over the hand-crafted unsupervised methods in the present setup. The supervised method is assumed to automatically learn features and combine multiple processes that are used in the unsupervised methods. Moreover the LSTM network is possibly able to learn some tricks that have not been discovered by human researchers. 
The two networks, i.e. LSTM-9 and LSTM-12, perform similarly for both the first nine noise types and the last three noise types, which indicates a good ability of such network to generalize to unseen noise types. The proposed method aims at learning a strategy that discriminates noise and speech frames mainly based on the stationarity of the magnitude sequence of a very limited set of frequencies (here 3 bins), rather than the wideband spectral structure of either speech or noise.  Therefore, the difference between the wideband spectral structure of the learning and test data does not impact the network generalization. However, we should mention that the proposed network cannot generalize to the extremely non-stationary noise, such as the machinegun noise in NOISEX92. 

\subsubsection{Noise PSD Estimation Example}

Fig. \ref{fig:example} shows an example of noise PSD estimation for a period of factory1 noise. Note that the result obtained with LSTM-9 is similar to the one obtained with LSTM-12, thus it is not shown. It can be seen that the LSTM-based estimator behaves similarly with the MMSE estimator in the sense that they both update the noise estimation smoothly at each frame. The main advantage of the LSTM-based estimator shown in this example is that it is sometimes able to track the abruptly increasing noise power.

\subsubsection{Speech Enhancement Results}
Fig. \ref{fig:se} shows the speech enhancement scores averaged over all the twelve noise types.  It is seen that all the three methods largely improve the performance measures over the noisy signal. Compared to the two unsupervised methods, the proposed method achieves  larger performance improvement by improving the accuracy of noise PSD estimation. However, the superiority of the proposed method for speech enhancement is not as prominent as the one for noise PSD estimation.

\section{Conclusion}
\label{sec4}

In this paper, we have proposed a noise PSD estimation method based on a supervised training of an LSTM network. The unsupervised methods \cite{martin2001,cohen2003,rangachari2006,li2016icassp,hendriks2010,gerkmann2012} previously demonstrated that an STFT magnitude sequence at one frequency bin contains rich information for noise PSD estimation. Our experiments show that an LSTM-based network is able to automatically exploit this information, and outperforms the unsupervised methods. 
Meanwhile, the proposed method preserves the merits of the unsupervised methods, namely generalizing well to the unseen speech/noise conditions.



\bibliographystyle{ieeetr}
\balance

\end{document}